\documentclass[10pt]{iopart}
\usepackage[dvipdfmx]{graphicx}
\usepackage{ulem}
\def\vector#1{\mbox{\boldmath $#1$}}

\begin{document}

\title{Solid--fluid transition of two- or three-dimensional systems with infinite-range interaction }
\author{Hisato Komatsu}
\address{ Graduate school of Arts and Sciences, The University of Tokyo, 3-8-1 Komaba, Meguro-ku, Tokyo 153-8902}
\ead{komatsu@huku.c.u-tokyo.ac.jp}
\begin{abstract}
It is difficult to derive the solid--fluid transition from microscopic models. We introduce particle systems whose potentials do not decay with distance and calculate their partition function exactly using a method similar to that for lattice systems with  infinite-range interaction. In particular, we investigate the behaviors of examples among these models, which become a triangular, body-centered cubic, face-centered cubic, or simple cubic lattice in low-temperature phase. The transitions of the first three examples are of the first order, and that of the last example is of the second order. Note that we define the solid phase as that whose order parameter, or Fourier component of the density, becomes nonzero, and the models we considered obey the ideal-gas law even in the solid phase.
\end{abstract}

\section{Introduction \label{intro}}
A fundamental problem in condensed matter physics and statistical physics is the theoretical derivation of solid--fluid phase transitions from microscopic Hamiltonians. There are several ways to define the solid and fluid phase. Some researches distinguished two phases from each other by the existence of the shear modulus\cite{MBorn}. Other researches considered the solid phase as the state with non-uniform periodic density. This definition is originated back to the theories by Vlasov\cite{Bazarov}, and also Kirkwood-Monroe\cite{KirMon}.
 In the standpoint of modern statistical physics, the latter definition is interpreted as the continuous translational symmetry breaking and the emergence of the corresponding order parameters. In this paper, we adopt the latter in order to focus on the symmetry breaking and the order parameters.
 Although many efforts have been made to develop the theory of solid--fluid transitions, most of them are based on phenomenological theory, and some do not take the translational symmetry breaking into account explicitly\cite{LJD,MOI,KirMon}. Only a few theoretical approaches, including density functional theory, have successfully overcome these problems to some extent \cite{RY,CA,DA}.
 Hence, to more deeply understand solid--fluid transitions, it is meaningful to search for and consider exactly solvable models that exhibit solid--fluid transitions. One of a few examples in which the solid--fluid transition is derived exactly is the research of Carmesin and Fan \cite{CF}. They considered a one-dimensional model with a cosine potential and hard core potential, and used the Hubbard--Stratonovich transformation for the cosine potential, as in the infinite-range XY model. However, their research used a special method to treat the hard core potential and was difficult to apply to more general cases. Therefore, it is natural for us to try to simplify their model by removing the hard core potential and to extend the discussion to higher-dimensional cases.

 In this paper, we propose higher-dimensional models whose particles interact with each other only by the summation of the cosine potentials and derive the solid--fluid transition exactly. First, we introduce the models in section 2, and then we calculate the partition function in sections 3 and 4. We next investigate the behavior of the order parameter and the order of the phase transition in section 5, and finally summarize this paper in section 6.

\section{Models \label{models}} 
We consider a $d$-dimensional model  depending on the values of $ \vector{k}_{\alpha }$, or the smallest reciprocal lattice vectors of the crystal under our consideration:
\begin{equation}
 H_{ \left\{ \vector{k}_{\alpha} \right\} } = \sum _{i} \frac{\vector{p}_i ^2}{2m} - \frac{J}{N} \sum _{\alpha }  \sum _{i,j} \cos \vector{k}_{\alpha } \cdot ( \vector{x}_i - \vector{x}_j ) - \sum _{\alpha } h_{\alpha} \sum _i \cos \vector{k}_{\alpha } \cdot \vector{x}_i ,
 \label{Hamiltonian1}
\end{equation}
 where $N$ is the number of particles; $\vector{x}_i$ and $\vector{p}_i$ are the position and momentum of the $i$th particle, respectively. We treat the spontaneous symmetry breaking by using the method of Bogoliubov quasi-averages, namely weak external field $h_{\alpha}$ with positive values is introduced to break symmetry. The set of $ \vector{k}_{\alpha }$ values contains all the reciprocal lattice vectors with the smallest sizes. When we introduce the $ \vector{k}_{\alpha }$ values, we regard a pair of reciprocal lattice vectors that are $-$1 times each other as one vector. For example, the set of $ \vector{k}_{\alpha }$ values is given as
\begin{equation}
 \vector{k}_1 = k \left(
 \begin{array}{c}
 1 \\
 0 \\
 \end{array}
 \right) , 
  \vector{k}_2 = k \left(
 \begin{array}{c}
 -1/2 \\
\sqrt{3}/2 \\
 \end{array}
 \right) , 
  \vector{k}_3 = k \left(
 \begin{array}{c}
 -1/2 \\
- \sqrt{3}/2 \\
 \end{array}
 \right)
 \label{k_tl}
\end{equation}
for a triangular lattice,
\begin{eqnarray}
  \vector{k}_1 & = & \frac{k}{\sqrt{2}} \left(
 \begin{array}{c}
 1 \\
 1 \\
 0 \\
 \end{array}
 \right) , 
  \vector{k}_2 = \frac{k}{\sqrt{2}} \left(
 \begin{array}{c}
 1 \\
 -1 \\
 0 \\
 \end{array}
 \right) , 
  \vector{k}_3 = \frac{k}{\sqrt{2}} \left(
 \begin{array}{c}
 0 \\
 1 \\
 1 \\
 \end{array}
 \right) , \nonumber \\ 
  \vector{k}_4 & = & \frac{k}{\sqrt{2}} \left(
 \begin{array}{c}
 0 \\
 1 \\
 -1 \\
 \end{array}
 \right) ,
   \vector{k}_5 = \frac{k}{\sqrt{2}} \left(
 \begin{array}{c}
 1 \\
 0 \\
 1 \\
 \end{array}
 \right) ,
   \vector{k}_6 = \frac{k}{\sqrt{2}} \left(
 \begin{array}{c}
 -1 \\
 0 \\
 1 \\
 \end{array}
 \right)
 \label{k_bcc}
\end{eqnarray}
for a body-centered cubic (bcc) lattice,
\begin{equation}
  \vector{k}_1 = \frac{k}{\sqrt{3}} \left(
 \begin{array}{c}
 1 \\
 1 \\
 1 \\
 \end{array}
 \right) , 
  \vector{k}_2 = \frac{k}{\sqrt{3}} \left(
 \begin{array}{c}
 1 \\
 -1 \\
 -1 \\
 \end{array}
 \right) , 
  \vector{k}_3 = \frac{k}{\sqrt{3}} \left(
 \begin{array}{c}
 -1 \\
 1 \\
 -1 \\
 \end{array}
 \right), 
  \vector{k}_4 = \frac{k}{\sqrt{3}} \left(
 \begin{array}{c}
 -1 \\
 -1 \\
 1 \\
 \end{array}
 \right)
 \label{k_fcc}
\end{equation}
for a face-centered cubic (fcc) lattice, and
\begin{equation}
  \vector{k}_1 = k \left(
 \begin{array}{c}
 1 \\
 0 \\
 \cdot \\
 \cdot \\
 \cdot \\
 0 \\
 \end{array}
 \right) , 
  \vector{k}_2 = k \left(
 \begin{array}{c}
 0 \\
 1 \\
 \cdot \\
 \cdot \\
 \cdot \\
 0 \\
 \end{array}
 \right) , ... ,
  \vector{k}_d = k \left(
 \begin{array}{c}
 0 \\
 0 \\
 \cdot \\
 \cdot \\
 \cdot \\
 1 \\
 \end{array}
 \right) 
 \label{k_sc}
\end{equation}
for the $d$-dimensional simple cubic lattice. Here, $k$ is the length of $\vector{k} _{\alpha}$. Notice that we introduce Hamiltonians which are different from each other for each crystals we consider. We consider crystals which include only one position for particles to rest in their primitive cells and span their reciprocal lattice space by $\vector{k} _{\alpha}$. Hence, we do not consider crystals such as hexagonal lattices and hexagonal close-packed lattices, which have several positions for particles to rest.

\section{Calculation of the partition function \label{calculation}}
In this section, we calculate the partition function of these models:
\begin{eqnarray}
Z_{ \left\{ \vector{k}_{\alpha} \right\} } & \equiv & \frac{1}{N!} \int  _{\Omega } \prod _i d \vector{x} _i  \int \prod _i d \vector{p} _i \exp \left( - \beta  H_{ \left\{ \vector{k}_{\alpha} \right\} } \right) \nonumber \\
 & = & \frac{1}{\Lambda ^{dN} N!} \int  _{\Omega } \prod _i d \vector{x} _i  \nonumber \\ 
 & & \cdot \exp \left( \frac{\beta J}{N} \sum _{\alpha } \sum _{i,j} \cos \vector{k} _{\alpha } \cdot ( \vector{x} _i - \vector{x} _j ) + \beta \sum _{\alpha } h _{\alpha } \sum _i \cos \vector{k} _{\alpha } \cdot \vector{x} _i \right),
\label{Z0}
\end{eqnarray}
\begin{eqnarray}
\mathrm{where} \ \ \frac{1}{\Lambda} = \sqrt{\frac{2 \pi m}{\beta }} .
\end{eqnarray}
 First, the cosine potential is transformed into one-body potentials with auxiliary variables $q_{c \alpha} , q_{s \alpha} $ using the Hubbard--Stratonovich transformation:
\begin{eqnarray}
 & & \exp \left( \frac{\beta J}{N} \sum _{\alpha } \sum _{i,j} \cos \vector{k} _{\alpha } \cdot ( \vector{x} _i - \vector{x} _j ) + \beta \sum _{\alpha } h _{\alpha } \sum _i \cos \vector{k} _{\alpha } \cdot \vector{x} _i \right) \nonumber \\
& = & \exp \left[ \sum _{\alpha } \left\{ \frac{\beta J}{N} \left( \sum _i \cos \vector{k} _{\alpha } \cdot \vector{x} _i \right) ^2 + \frac{\beta J}{N} \left( \sum _i \sin \vector{k} _{\alpha } \cdot \vector{x} _i \right) ^2 + \beta h_{\alpha} \sum _i \cos \vector{k} _{\alpha } \cdot \vector{x} _i \right\} \right] \nonumber \\
& = & \int \prod _{\alpha } \left( \frac{\beta J N}{\pi} dq_{c \alpha } dq_{s \alpha }  \right) \exp \left[ \sum _{\alpha } \left\{ - \beta J N ( q _{c \alpha } ^2 +q _{s \alpha } ^2 ) \right. \right. \nonumber \\
& & \left. \left. + \beta ( 2 J q_{c \alpha } + h_{\alpha }) \sum _i \cos \vector{k} _{\alpha } \cdot \vector{x} _i  + 2 \beta J q_{s \alpha } \sum _i \sin \vector{k} _{\alpha } \cdot \vector{x} _i \right\} \right]  .
\end{eqnarray}
Then, the partition function is expressed as
\begin{eqnarray}
Z_{ \left\{ \vector{k}_{\alpha} \right\} } & = & \frac{1}{\Lambda ^{dN} N!} \int \prod _{\alpha } \left( \frac{\beta J N}{\pi} dq_{c \alpha } dq_{s \alpha }  \right) \exp \left[  \sum _{\alpha } - \beta J N ( q _{c \alpha } ^2 +q _{s \alpha } ^2 ) \right] \int  _{\Omega } \prod _i d \vector{x} _i \nonumber \\
& & \cdot \exp \left[ \sum _{\alpha } \left\{ \beta ( 2 J q_{c \alpha } + h_{\alpha }) \sum _i \cos \vector{k} _{\alpha } \cdot \vector{x} _i  + 2 \beta J q_{s \alpha } \sum _i \sin \vector{k} _{\alpha } \cdot \vector{x} _i \right\} \right] \nonumber \\
& \simeq & \frac{1}{\Lambda ^{dN} N!} \left[ \max _{q _{c \alpha} , q _{s \alpha} } \left[ \exp \left( - \sum _{\alpha} \beta J (q _{c \alpha} ^2 + q _{s \alpha} ^2 ) \right) \int _{\Omega } d \vector{x} \right. \right. \nonumber \\
& & \left. \left. \cdot \exp \left\{ \sum _{\alpha } \left\{ \beta ( 2 J q_{c \alpha } + h_{\alpha }) \cos \vector{k} _{\alpha } \cdot \vector{x}  + 2 \beta J q_{s \alpha } \sin \vector{k} _{\alpha } \cdot \vector{x}  \right\} \right\} \right] \right] ^N .
\label{Z1}
\end{eqnarray}
This method of calculating the partition function is similar to that for the infinite-range XY model and that used in previous works for one-dimensional systems \cite{FP,CF}.
We next maximize the right-hand side of equation (\ref{Z1}) under the condition that the values of $r_{\alpha } \equiv \sqrt{ q _{c \alpha} ^2 + q _{s \alpha} ^2 }$ are fixed.
 The maximum occurs when $q_{c \alpha} = r_{\alpha } , q_{s \alpha} =0$ because of the weak external fields $\left\{ h_{\alpha} \right\} $. The validity of this analysis is discussed in the Appendix. Considering this discussion, the partition function is expressed as the maximum of an $r_{\alpha } $ function:
\begin{eqnarray}
Z_{ \left\{ \vector{k}_{\alpha} \right\} } & \simeq & \frac{1}{\Lambda ^{dN} N!} \left[ \max _{r _{\alpha} >0 } \left[ \exp \left( - \sum _{\alpha} \beta J r _{\alpha} ^2 \right) \right. \right. \nonumber \\
& & \left. \left. \cdot \int _{\Omega } d \vector{x} \cdot \exp \left\{ \sum _{\alpha } \left\{ \beta ( 2 J r_{\alpha } + h_{\alpha }) \cos \vector{k} _{\alpha } \cdot \vector{x}  \right\} \right\} \right] \right] ^N \nonumber \\
 & = & \frac{1}{\Lambda ^{dN} N!} \left[ \max _{r _{\alpha} >0 } \left[ \exp \left( - \sum _{\alpha} \beta J \left( r _{\alpha} -\frac{h_{\alpha } }{2J} \right) ^2 \right) \right. \right. \nonumber \\
& & \left. \left. \cdot \int _{\Omega } d \vector{x} \cdot \exp \left\{ \sum _{\alpha } \left\{2 \beta J r_{\alpha }  \cos \vector{k} _{\alpha } \cdot \vector{x}  \right\} \right\} \right] \right] ^N .
\label{Z1.5}
\end{eqnarray}
 Taking the periodicity of the potential into account, we express the integral in equation (\ref{Z1.5}) as 
\begin{equation}
\int _{\Omega } d \vector{x}  \cdot \exp \left\{ \sum _{\alpha } \left\{2 \beta J r_{\alpha }  \cos \vector{k} _{\alpha } \cdot \vector{x} \right\} \right\} = \frac{V}{V_{\mathrm{p} } } \Phi _{\mathrm{p} } \left( \left\{ r_{\alpha} \right\} , \left\{ \vector{k}_{\alpha} \right\}  \right) ,
\end{equation}
 where $ V_{\mathrm{p}}$ and $ \Phi _{\mathrm{p}}$ represent the integrals over a primitive cell of the supposed crystal and are given as
\begin{eqnarray}
 \Phi _{\mathrm{p} } \left( \left\{ r_{\alpha} \right\} , \left\{ \vector{k}_{\alpha} \right\}  \right) & \equiv & \int _{\mathrm{primitive \ cell} } d \vector{x}  \cdot \exp \left\{ \sum _{\alpha } \left\{2 \beta J r_{\alpha } \cos \vector{k} _{\alpha } \cdot \vector{x} \right\} \right\} \label{Kpr} , \\
 V_{\mathrm{p} }& \equiv & \int _{\mathrm{primitive \ cell} } d \vector{x}  \cdot 1 \label{Vpr} .
\end{eqnarray}
Hence, the partition function is expressed as
\begin{equation}
 Z_{ \left\{ \vector{k}_{\alpha} \right\} } = \frac{V^N}{\Lambda ^{dN} N!} \left[ \frac{1}{V_{\mathrm{p} } } \max _{r _{\alpha} >0 } \left[ \exp \left( - \sum _{\alpha} \beta J \left( r _{\alpha} -\frac{h_{\alpha } }{2J} \right) ^2 \right) \Phi _{\mathrm{p} } \left( \left\{ r_{\alpha} \right\} , \left\{ \vector{k}_{\alpha} \right\} \right) \right] \right] ^N .
 \label{Z2}
\end{equation}
 The solid--fluid transition is investigated using this equation. At the end of this section, we discuss the thermodynamic properties of this system. The order parameters, or the Fourier components of the density, are given as
\begin{equation}
 \left< \rho _{\vector{k} _{\alpha} } \right> \equiv \left< \frac{1}{N} \sum _{i} \cos \vector{k} _{\alpha} \cdot \vector{x} _i  \right> = \left. \frac{1}{\beta N} \frac{\partial}{\partial h_{\alpha}} \log Z \right| _{ h_{\alpha} \rightarrow +0} = \lim _{ h_{\alpha} \rightarrow +0} \tilde{r}_{\alpha} \left( \left\{ h_{\alpha} \right\} \right),
 \label{op}
\end{equation}
where $\tilde{r}_{\alpha} \left( \left\{ h_{\alpha} \right\} \right)$ represents the value of $r _{\alpha}$ that yields the maximum in equation (\ref{Z2}). The equation of state is derived from the partition function as
\begin{equation}
 P  = \frac{\partial}{\partial V} \left( \frac{\log Z}{\beta} \right) = \frac{N}{\beta V} .
\end{equation}
This is nothing but the ideal-gas law, which holds even in the solid phase. This strange property results from the fact that although the cosine potentials tend to restrict the positions of the particles to a set of lattice points, it does not matter on which of these points particles rest. 

\section{Analytic expression of the partition function \label{Bessel}}
 In this section, we discuss an analytical expression of $\Phi _{\mathrm{p} } $ starting from equation (\ref{Z1.5}). First, the exponential in the integral is expanded as follows:
\begin{eqnarray}
 &  & \int _{\Omega } d \vector{x} \cdot \exp \left\{ \sum _{\alpha } \left\{2 \beta J r_{\alpha }  \cos \vector{k} _{\alpha } \cdot \vector{x}  \right\} \right\}  \nonumber \\
   & = &  \int _{\Omega } d \vector{x} \cdot \prod _{\alpha} \exp \left\{ \beta J r_{\alpha }  e^{i \vector{k} _{\alpha } \cdot \vector{x} } \right\} \exp \left\{ \beta J r_{\alpha } e^{-i \vector{k} _{\alpha } \cdot \vector{x} } \right\} \nonumber \\
   & = & \int _{\Omega } d \vector{x} \cdot \prod _{\alpha}  \left\{ \sum _{n _{\alpha}, n' _{\alpha} } \frac{ \left( \beta J r_{\alpha } \right) ^{n _{\alpha } } e^{i n _{\alpha }  \vector{k} _{\alpha } \cdot \vector{x} } }{n _{\alpha} !} \cdot \frac{ \left( \beta J r_{\alpha } \right) ^{n' _{\alpha } } e^{-i n' _{\alpha } \vector{k} _{\alpha } \cdot \vector{x} } }{n' _{\alpha} !} \right\}  \nonumber \\
   & = & \sum _{\vector{n}, \vector{n'}} \left\{ \prod _{\alpha} \left( \frac{ \left( \beta J r_{\alpha } \right) ^{n _{\alpha } + n' _{\alpha} } }{n _{\alpha} ! n' _{\alpha}!} \right) \cdot \int _{\Omega } d \vector{x} \exp \left\{i \sum _{\alpha} ( n _{\alpha } - n' _{\alpha })  \vector{k} _{\alpha } \cdot \vector{x} \right\} \right\} .
\label{Z3}
\end{eqnarray}
We ignore the values of the weak external field $h_{\alpha}$ in this section because their role ended when they gave the value of the order parameter under the condition that $h_{\alpha} \rightarrow +0$ in equation (\ref{op}). The integral on the right-hand side of equation (\ref{Z3}) is calculated as
\begin{equation}
  \int _{\Omega } d \vector{x} \exp \left\{i \sum _{\alpha} ( n _{\alpha } - n' _{\alpha })  \vector{k} _{\alpha } \cdot \vector{x} \right\} = \left\{
\begin{array}{c}
V \ \ \mathrm{if} \ \  \sum _{\alpha} ( n _{\alpha } - n' _{\alpha })  \vector{k} _{\alpha } =0, \\ 
\\
O(1) \ll V \ \ \ \mathrm{otherwise}. \\ 
\end{array}
\right. 
\end{equation}
Consequently, the contribution of the terms that have nonzero values of $ \sum _{\alpha} ( n _{\alpha } - n' _{\alpha })  \vector{k} _{\alpha } $ is sufficiently small. Hence, the partition function is given as
\begin{eqnarray}
Z_{ \left\{ \vector{k}_{\alpha} \right\} } & = & \frac{V^N}{\Lambda ^{dN} N!} \left[ \max _{r _{\alpha} >0 } \left[ \exp \left( - \sum _{\alpha} \beta J r _{\alpha} ^2 \right) \sum _{\vector{n}, \vector{n'}}   \right. \right. \nonumber \\
& & \left. \left. \cdot \left\{ \prod _{\alpha} \left( \frac{ \left( \beta J r_{\alpha } \right) ^{n _{\alpha } + n' _{\alpha} } }{n _{\alpha} ! n' _{\alpha}!} \right) \delta _{ \sum _{\alpha} ( n _{\alpha } - n' _{\alpha })  \vector{k} _{\alpha } , 0} \right\} \right] \right] ^N .
\label{Z4}
\end{eqnarray}
This is a general expression of the partition function, independent of the set of $\vector{k}_{\alpha}$ values.

In the following, the partition function for each lattice structure is discussed. 
For the triangular lattice, taking the fact that $\vector{k} _1 + \vector{k} _2 + \vector{k} _3 = 0$ into account, we simplify equation (\ref{Z4}) as
\begin{eqnarray}
Z_{\mathrm{tl}} & = & \frac{V^N}{\Lambda ^{2N} N!} \left[ \max _{r _{\alpha} >0 } \left[ \exp \left( - \sum _{\alpha} \beta J r _{\alpha} ^2 \right) \sum _{a = -\infty} ^{\infty} \sum _{\vector{n}}  \left\{ \prod _{\alpha} \left( \frac{ \left( \beta J r_{\alpha } \right) ^{2n _{\alpha } + |a| } }{n _{\alpha} ! (n _{\alpha} + |a| )!} \right) \right\} \right] \right] ^N \nonumber \\
   & = & \frac{V^N}{\Lambda ^{2N} N!} \left[ \max _{r _{\alpha} >0 } \left[ \exp \left( - \sum _{\alpha} \beta J r _{\alpha} ^2 \right) \sum _{a = -\infty} ^{\infty} \prod _{\alpha } \left\{ I _{|a|} (2 \beta J r_{\alpha }) \right\}  \right] \right] ^N ,
\label{Z_TL0}
\end{eqnarray}
where $I _{\nu}$ is the $\nu$th modified Bessel function:
\begin{equation}
I_{\nu} (x) = \sum _n \frac{1}{n! (n+\nu )!} \left( \frac{x}{2} \right) ^{2n + \nu} .
\label{Modified_Bessel}
\end{equation}
Assuming that this system does not break the symmetry under the permutation of $\alpha$, and hence all $r_{\alpha }$ have the same value $r$, we obtain
\begin{equation}
Z_{\mathrm{tl}}  =  \frac{V^N}{\Lambda ^{2N} N!} \left[ \max _{r >0 } \left[ \exp \left( - 3 \beta J r ^2 \right) \sum _{a = -\infty} ^{\infty} \left\{ I _{|a|} (2 \beta J r) \right\} ^3 \right] \right] ^N .
\label{Z_TL}
\end{equation}
 For an fcc lattice, using the fact that the relation $\vector{k} _1 + \vector{k} _2 + \vector{k} _3 + \vector{k} _4 = 0$ exists, we obtain
\begin{equation}
Z_{\mathrm{fcc}}  =  \frac{V^N}{\Lambda ^{3N} N!} \left[ \max _{r >0 } \left[ \exp \left( - 4 \beta J r ^2 \right) \sum _{a = -\infty} ^{\infty} \left\{ I _{|a|} (2 \beta J r) \right\} ^4 \right] \right] ^N .
\label{Z_fcc}
\end{equation}
Although the case of the bcc lattice is more complicated, a similar expression for the partition function appears after some calculations:
\begin{eqnarray}
Z_{\mathrm{bcc}}  & = & \frac{V^N}{\Lambda ^{3N} N!} \left[ \max _{r >0 } \left[ \exp \left( - 6 \beta J r ^2 \right)  \right. \right. \nonumber \\
& & \sum _{a,b,c = -\infty} ^{\infty} \cdot \left. \left. I _{|a|} (2 \beta J r) I _{|b|} (2 \beta J r) I _{|c|} (2 \beta J r) \tilde{I} _{a,b} (2 \beta J r) \tilde{I} _{b,c} (2 \beta J r) \tilde{I} _{c,a} (2 \beta J r) \right] \right] ^N ,
\label{Z_bcc} 
\end{eqnarray}
where
\begin{equation}
  \tilde{I} _{\mu , \nu} (x)  = \left\{
\begin{array}{c}
 \sum _n \frac{1}{(n+| \mu | )! (n+ | \nu | )!} \left( \frac{x}{2} \right) ^{2n + | \mu |+ | \nu |} \ \ \mathrm{if} \ \ \mathrm{sgn} ( \mu ) \neq \mathrm{sgn} ( \nu ) , \\ 
I _{|a|+|b|} (x) \ \ \mathrm{if} \ \ \mathrm{sgn} ( \mu ) = \mathrm{sgn} ( \nu ) . \\ 
\end{array}
\right.
\end{equation}
For a simple cubic lattice, because the smallest reciprocal lattice vectors $ \vector{k} _1 , \vector{k} _2 ,  ... ,\vector{k} _d $ are linearly independent, the nonzero terms in equation (\ref{Z4}) fulfill $n_{\alpha} = n' _{\alpha}$ for all $\alpha$ values. Hence, the partition function is written as
\begin{equation}
Z_{\mathrm{sc}}  =  \frac{V^N}{\Lambda ^{dN} N!} \left[ \max _{r >0 } \left[ \exp \left( - \beta J r ^2 \right) I _0  (2 \beta J r) \right] \right] ^{dN} .
\label{Z_sc}
\end{equation}
 This is proportional to the $d$th power of the partition function of a one-dimensional system with the cosine potential.

\section{Behavior of the phase transition}
\subsection{Temperature dependence of the order parameter \label{order}}
\begin{figure}[!hbp]
\begin{center}
\includegraphics[width = 8.0cm]{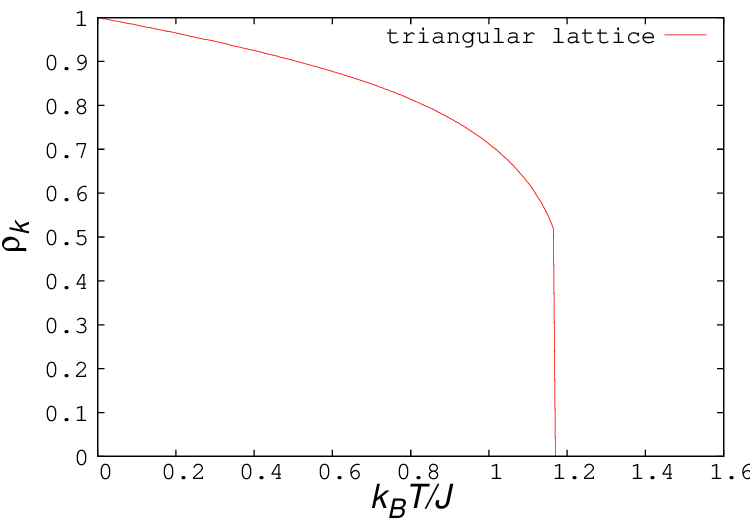}
\caption{Temperature dependence of the order parameter for the triangular lattice model}
\label{F_TL}
\end{center}
\end{figure}

\begin{figure}[!hbp]
\begin{center}
\includegraphics[width = 8.0cm]{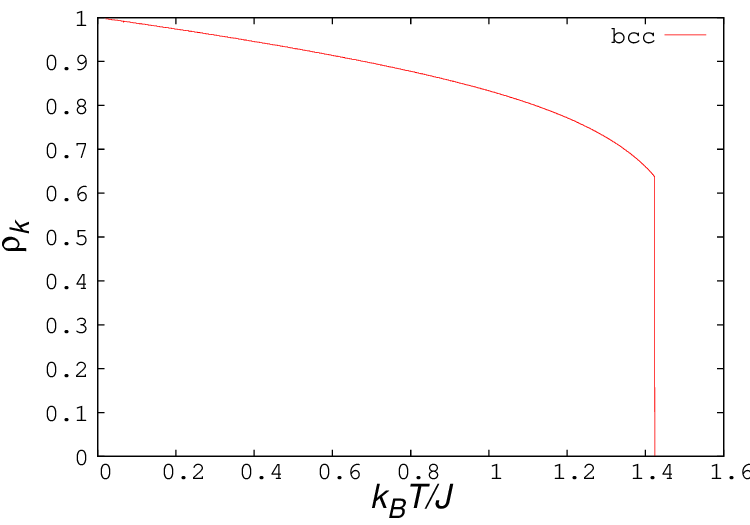}
\caption{Temperature dependence of the order parameter for the bcc lattice model}
\label{F_bcc}
\end{center}
\end{figure}

\begin{figure}[!hbp]
\begin{center}
\includegraphics[width = 8.0cm]{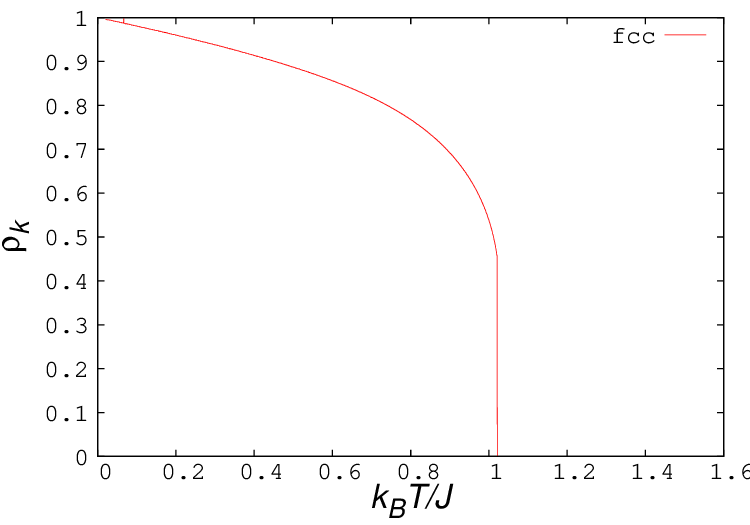}
\caption{Temperature dependence of the order parameter for the fcc lattice model}
\label{F_fcc}
\end{center}
\end{figure}

\begin{figure}[!hbp]
\begin{center}
\includegraphics[width = 8.0cm]{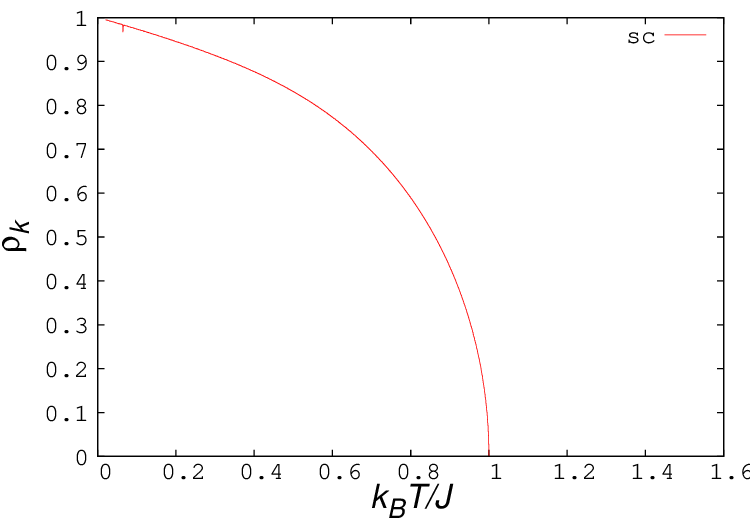}
\caption{Temperature dependence of the order parameter for the simple cubic lattice model}
\label{F_sc}
\end{center}
\end{figure}

\begin{figure}[!hbp]
\begin{center}
\includegraphics[width = 8.0cm]{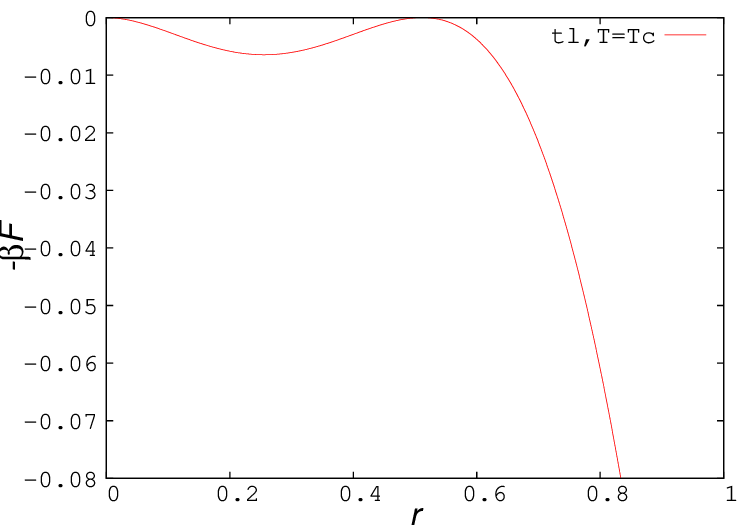}
\caption{$r$ dependence of $-\beta F$ for the triangular lattice model at transition point without difference in constant}
\label{Free_tl}
\end{center}
\end{figure}

In this section, we investigate the behavior of the order parameter $r_{\alpha}$ by calculating equation (\ref{Z2}) numerically. If we examine equation (\ref{Z4}) in the previous section carefully, we find that this partition function is independent of $k$, or the length of $\vector{k}_{\alpha}$. Hence, we let $k=1$ and also assume that all $r_{\alpha}$ have the same value as in the previous section. The results are shown in figures \ref{F_TL}--\ref{F_sc}.
To identify the order of the transition, we investigated the $r$ dependence of the logarithm of $\exp \left( -n' \beta J r^2 \right) \Phi_{\mathrm{p}} /V_{\mathrm{p}}$, which appears on the right-hand side of equation (\ref{Z2}). This function corresponds to $- \beta$ times the Landau free energy except for the difference in the constant. For the triangular lattice, this function has two peaks, the heights of which are interchanged at the transition point, as we show in figure \ref{Free_tl}. Hence, the solid--fluid transition is a first-order transition in this case. The transition of bcc and fcc lattices is also found to be a first-order transition by a similar investigation. However, the transition of the simple cubic lattice is of the second order because the behavior of the order parameter is the same as in the infinite-range XY model. 
Note that although the power series expansions of the partition function of the fcc lattice model do not include odd-order terms of the order parameter, the transition in this model is of the first order. This is caused by the sixth- or higher-order terms. 

\clearpage
\subsection{Effect of additional potential with higher wavenumber on the simple cubic lattice \label{remark}}
\begin{figure}[!hbp]
\begin{center}
\includegraphics[width = 8.0cm]{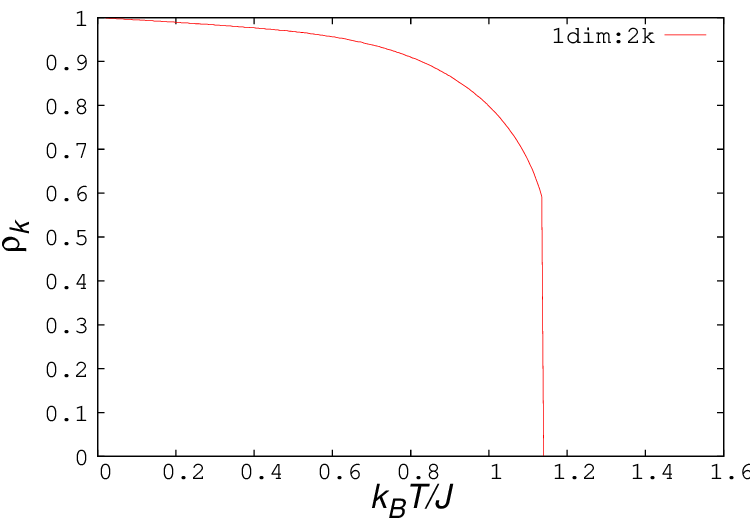}
\caption{Temperature dependence of the order parameter for equation (\ref{Hamiltonian2k})}
\label{F_1d2k}
\end{center}
\end{figure}

As we saw in section \ref{order}, the model of the simple cubic lattice has a second-order transition if $\left\{ \vector{k} _{\alpha} \right\}$ includes only the smallest reciprocal lattice vectors regardless of the spatial dimension $d$. However, it is possible that the transition is of the first order if $\left\{ \vector{k} _{\alpha} \right\}$ includes other reciprocal lattice vectors.
We consider the case of $d=1$ that contains $k$ and $2k$ in $\left\{ \vector{k} _{\alpha} \right\}$ as an example:
\begin{equation}
 H_{ \left\{ \vector{k}_{\alpha} \right\} } = \sum _{i} \frac{p_i ^2}{2m} - \frac{J}{N}  \sum _{i,j} \left\{ \cos k ( x_i -x_j ) + \cos 2k ( x_i -x_j ) \right\} -  h \sum _i \cos k x_i .
 \label{Hamiltonian2k}
\end{equation}
The partition function of this model is calculated by a method similar to that in sections \ref{calculation} and \ref{Bessel} as
\begin{eqnarray}
Z_{ \left\{ \vector{k}_{\alpha} \right\} } & = & \frac{V^N}{\Lambda ^{N} N!} \left[ \max _{r _{\alpha} >0 } \left[ \exp \left( - \sum _{\alpha =1,2} \beta J r _{\alpha} ^2 \right) \sum _{\vector{n}, \vector{n'}} \right. \right. \nonumber \\
& & \left. \left. \cdot \left\{ \prod _{\alpha = 1,2} \left( \frac{ \left( \beta J r_{\alpha } \right) ^{n _{\alpha } + n' _{\alpha} } }{n _{\alpha} ! n' _{\alpha}!} \right)
 \delta _{ \sum _{\alpha} \alpha \cdot ( n _{\alpha } - n' _{\alpha }) , 0} \right\} \right] \right] ^N  \nonumber \\
& = & \frac{V^N}{\Lambda ^{N} N!} \left[ \max _{r _{\alpha} >0 } \left[ \exp \left( - \sum _{\alpha =1,2} \beta J r _{\alpha} ^2 \right) \sum _{\vector{n}}  \sum _{a = -\infty} ^{\infty} \right. \right. \nonumber \\ 
& & \left. \left. \cdot \left( \frac{ \left( \beta J r_{1 } \right) ^{2n _{1 } + 2|a| } \left( \beta J r_{2 } \right) ^{2n _{2 } + |a| } }{n _{1} ! (n _{1}+2|a|)! n_{2} ! (n_{2} +|a|)! } \right) \right] \right] ^N  \nonumber \\
& = & \frac{V^N}{\Lambda ^{N} N!} \left[ \max _{r _{\alpha} >0 } \left[ \exp \left( - \sum _{\alpha =1,2} \beta J r _{\alpha} ^2 \right) \left\{ \sum _{a = -\infty} ^{\infty}   I _{2|a|} (2 \beta J r_1 ) I _{|a|} (2 \beta J r_2 ) \right\}  \right] \right] ^N .
\label{Z1d2k}
\end{eqnarray}
 Consequently, the solid--fluid transition in this model is of the first order. The temperature dependence of the order parameter is shown in figure \ref{F_1d2k}. 
If we ignore the terms with $a \neq 0$ in equation (\ref{Z1d2k}), the summation appearing on the right-hand side of this equation reduces to the simple product of the zeroth modified Bessel function, and the transition is judged to be of the second order, like that of equation (\ref{Z_sc}). In other words, the terms with $a \neq 0$ cause the transition to be of the first order.

\clearpage
\section{Summary}
We introduced two- or higher-dimensional models in which particles interact with each other by the summation of the cosine potentials and calculated their partition function. In particular, we investigated four examples that form a triangular, bcc, fcc, or simple cubic lattice below the transition point. The phase transitions of the first three examples are the first order, 
and that of the simple cubic lattice is a second-order transition. 

Because the potentials of these models do not decay with distance and are treated similarly to the infinite-range model of spin systems, these models are thought to be called ``mean-field models'' of  the solid--fluid transition.  Hence, these models are expected to be used to constitute the ``mean-field approximation'' of solid--fluid transitions of simple particle systems, such as the three-dimensional Lennard-Jones system. In order to construct the approximation properly and obtain a non-trivial equation of state, not the ideal-gas law, it is necessary to introduce an effective coupling between the order parameter of the solid phase  and the density or the volume of the system. 

\section*{Acknowledgements}
The author thanks to Koji Hukushima for helpful discussions. 

\appendix

\section{Proof of $q_{c \alpha } = r_{\alpha}, q_{s \alpha } = 0$ in section \ref{calculation} \label{app1} } 
To prove that the integral in equation (\ref{Z1}) reaches its maximum when $q_{c \alpha } = r_{\alpha}, q_{s \alpha } = 0$ under the condition that $r_{\alpha } \equiv \sqrt{ q _{c \alpha} ^2 + q _{s \alpha} ^2 }$ are fixed, we first expand the integral in equation (\ref{Z1}) in the same way as that in equation (\ref{Z3}).
\begin{eqnarray}
 &  & \int _{\Omega } d \vector{x} \cdot \exp  \left\{ \sum _{\alpha } \left\{ \beta ( 2 J q_{c \alpha } + h_{\alpha }) \cos \vector{k} _{\alpha } \cdot \vector{x}  + 2 \beta J q_{s \alpha } \sin \vector{k} _{\alpha } \cdot \vector{x}  \right\} \right\}  \nonumber \\
   & = &  \int _{\Omega } d \vector{x} \cdot \prod _{\alpha} \exp \left\{ \beta (J q_{c \alpha } + h_{\alpha}/2 - i J q_{s \alpha } ) e^{i \vector{k} _{\alpha } \cdot \vector{x} } \right\} \nonumber \\
 & & \cdot \exp \left\{ \beta (J q_{c \alpha } + h_{\alpha}/2 + i J q_{s \alpha } ) e^{-i \vector{k} _{\alpha } \cdot \vector{x} } \right\}  \nonumber \\
   & = & \int _{\Omega } d \vector{x} \cdot \prod _{\alpha}  \left\{ \sum _{n _{\alpha}, n' _{\alpha} } \frac{ \left( \beta J z_{\alpha } \right) ^{n _{\alpha } } e^{i n _{\alpha }  \vector{k} _{\alpha } \cdot \vector{x} } }{n _{\alpha} !} \cdot \frac{ \left( \beta J \bar{z}_{\alpha } \right) ^{n' _{\alpha } } e^{-i n' _{\alpha } \vector{k} _{\alpha } \cdot \vector{x} } }{n' _{\alpha} !} \right\}  \nonumber \\
   & = & \sum _{\vector{n}, \vector{n'}} \left\{ \prod _{\alpha} \left( \frac{ \left( \beta J z_{\alpha } \right) ^{n _{\alpha } }  \left( \beta J \bar{z}_{\alpha } \right) ^{n' _{\alpha} } }{n _{\alpha} ! n' _{\alpha}! } \right) \cdot \int _{\Omega } d \vector{x} \exp \left\{ i \sum _{\alpha} ( n _{\alpha } - n' _{\alpha })  \vector{k} _{\alpha } \cdot \vector{x} \right\}  \right\}  \nonumber \\
    & = & \sum _{\vector{n}, \vector{n'}} \left\{ \prod _{\alpha} \left( \frac{ \left( \beta J z_{\alpha } \right) ^{n _{\alpha } }  \left( \beta J \bar{z}_{\alpha } \right) ^{n' _{\alpha} } }{n _{\alpha} ! n' _{\alpha}! } \right) \cdot \delta _{ \sum _{\alpha} ( n _{\alpha } - n' _{\alpha })  \vector{k} _{\alpha } , 0} \right\} ,  
\label{ZA1}
\end{eqnarray}
where
\begin{equation}
 J z_{\alpha } \equiv J q_{c \alpha } + h_{\alpha}/2 - i J q_{s \alpha } .
\end{equation}
By substituting (\ref{ZA1}) into (\ref{Z1}), the partition function is given as 
\begin{eqnarray}
Z & = &  \frac{1}{\Lambda ^{dN} N!} \left[ \max _{q _{c \alpha} , q _{s \alpha} } \left[ \exp \left( - \sum _{\alpha} \beta J (q _{c \alpha} ^2 + q _{s \alpha} ^2 ) \right) \right. \right. \cdot \nonumber \\
& & \left. \left. \sum _{\vector{n}, \vector{n'}} \left\{ \prod _{\alpha} \left( \frac{ \left( \beta J z_{\alpha } \right) ^{n _{\alpha } }  \left( \beta J \bar{z}_{\alpha } \right) ^{n' _{\alpha} } }{n _{\alpha} ! n' _{\alpha}! } \right) \cdot \delta _{ \sum _{\alpha} ( n _{\alpha } - n' _{\alpha })  \vector{k} _{\alpha } , 0} \right\} \right] \right] ^N .
\label{ZA2}
\end{eqnarray}
The proposition we try to prove here will be fulfilled if the real part of each term appearing in the summation of equation (\ref{ZA2}) becomes maximum when $q_{c \alpha } = r_{\alpha}, q_{s \alpha } = 0$. We show this fact starting from the following inequality;
\begin{eqnarray}
\mathrm{Re} \left( \prod _{\alpha} z_{\alpha } ^{n _{\alpha }}  \bar{z}_{\alpha } ^{n' _{\alpha }} \right) & \leq & \left| z \right| ^{n _{\alpha } + n' _{\alpha }} = \sqrt{ \left( q _{c \alpha} + \frac{h_{\alpha} }{2J} \right) ^2 + q _{s \alpha} ^2} = \sqrt{ r _{ \alpha}^2 + 2 q _{c \alpha} \cdot \frac{h_{\alpha} }{2J} + \left( \frac{h_{\alpha} }{2J} \right) ^2 }  \nonumber \\
 & \leq &  r _{ \alpha} + \frac{h_{\alpha} }{2J} \ \ .
\label{zineq1}
\end{eqnarray}
The equality is satisfied when $q_{c \alpha } = r_{\alpha}, q_{s \alpha } = 0$; hence, the integral in equation (\ref{Z1}) takes its maximum in this case.
Furthermore, considering the term when $n_{\alpha '} = n' _{\alpha '} = 1$ for $\alpha = \alpha ' $ and  $n_{\alpha} = n' _{\alpha} = 0$ for other $\alpha$ values, the inequality (\ref{zineq1}) becomes
\begin{eqnarray}
\mathrm{Re} z_{\alpha '}  \bar{z}_{\alpha '} & = & \left| z \right| ^2 =  r _{ \alpha}^2 + 2 q _{c \alpha} \cdot \frac{h_{\alpha} }{2J} + \left( \frac{h_{\alpha} }{2J} \right) ^2 \nonumber \\
 & \leq & \left( r _{ \alpha} + \frac{h_{\alpha} }{2J} \right) ^2 \ \ .
\label{zineq2}
\end{eqnarray}
Both sides of the inequality (\ref{zineq2}) are equal to each other if and only if $q_{c \alpha } = r_{\alpha}, q_{s \alpha } = 0$. Because of this example, the case that $q_{c \alpha } = r_{\alpha}, q_{s \alpha } = 0$ turns out to be the unique one that yields the maximum value of the integral in equation (\ref{Z1}). 

\section*{reference}

\end{document}